# Coexisting polarization mechanisms in ferroelectric uniaxial tetragonal tungsten bronze $Ca_{0.3}Ba_{0.7}Nb_2O_6$ (CBN-30)

Elena Buixaderas[1], Šarūnas Svirskas[2], Christelle Kadlec[1], Maxim Savinov[1], Patricija Lapienytė[2], Anirudh K.R[1], Cosme Milesi-Brault[1], Dmitry Nuzhnyy[1], Jan Dec[3]

[1]Institute of Physics of the Czech Academy of Sciences, Na Slovance 2, 18221 Prague, Czech Republic
[2]Faculty of Physics, Vilnius University, Saulėtekio al. 9, LT-10222 Vilnius, Lithuania
[3]Institute of Materials Science, University of Silesia, Bankowa 12, PL 40-007 Katowice, Poland

**Abstract**

Using a broad band dielectric spectroscopy approach (1 to $10^{14}$ Hz) we prove that the tungsten bronze $Ca_{0.3}Ba_{0.7}Nb_2O_6$ (CBN-30) displays a ferroelectric phase transition of mixed displacive and order-disorder character, and its paraelectric phase does not show traces of relaxor behaviour but precursor effects as polar fluctuations below about 550 K. The analysis of the sub-MHz dielectric response together with infrared and Raman spectroscopy reveals that simultaneous polarization mechanisms are responsible for the phase transition. The comparison of the excitations found in CBN-30 with those of $(Sr,Ba)Nb_2O_6$ reveals that these mechanisms are congruous, although in CBN-30 the main relaxation process behaves differently due to the different domain structure.

The excitations are phenomenologically assigned to phonons, to an anharmonic vibration of cationic origin which plays the role of a soft central mode ($\nu_{THz}$), and to a relaxation in the GHz range ($\nu_{01}$) probably due to polarization fluctuations of nanometric size which carries the main part of the permittivity and splits below $T_C$ into several weaker excitations with different polarization correlation lengths. The overall dielectric response is therefore explained by the coexistence of several excitations with different thermal behaviors, corroborating the complexity of the tetragonal tungsten bronze structures.

**INTRODUCTION**

Besides the perovskites, tetragonal tungsten-bronzes (TTBs) also have a great potential to be explored in electronics and electro-optics, due to the peculiarities of their crystal structure. Their unit cell can be represented as $(A1,A2)_6C_4(B1_2,B2_8)O_{30}$. The crystallographic structure is built up by a network of corner-sharing oxygen octahedra, forming shaped channels along the tetragonal main axis (the *c*-axis) (A1- squared, A2- pentagonal, and C-triangular). Therefore, the TTB structure is less compact in comparison to the perovskite-like framework. Thus, these channels can be filled with different cations. Triangular channels are the smallest and can only be filled with small ions such as $Li^{1+}$, but in most cases they remain empty. Even in this case TTBs have four different sites that can be filled with distinct cations, leaving more possibilities to explore compared to the perovskite structure. When B1 and B2 atoms are the same, their primitive unit cell is 5 times smaller and it can be written as $(A1,A2)B_2O_6$, although the true unit cell has five of these units (Z=5). The main representative of the unfilled TTBs is $(Sr_xBa_{1-x})Nb_2O_6$ (SBN-X, X=100x), which has been extensively investigated due to its excellent nonlinear properties for electro-optics [1], and because it has one of the highest permittivity values among ferroelectric crystals [2]. In addition, it shows a perfect tuning ferroelectric-relaxor behaviour depending on the Sr amount present in the crystal lattice [3]. Compositions with Ca instead of Sr $(Ca_xBa_{1-x})Nb_2O_6$ (CBN-X), display even higher permittivity and phase transition temperature [4], making this material an excellent candidate for high-T applications.

Important advances in the understanding of the polarization distribution and the relaxor nature of TTBs have been made, using a broadband dielectric spectroscopy, diffuse scattering and computational simulations [5, 6, 7, 8]. The combination of random cationic substitution, vacancies and distortion of the oxygen octahedra produces a complicated incommensurate modulated structure which indirectly leads to the relaxor behaviour, although the exact mechanism is still under debate. The presence of two crystallographic sites for the Nb atom appears to be fundamental for the occurrence of the ferroelectricity [7], and the existence of some ferroelectric-antiferroelectric competition has also been recently reported [9].
Crystallographic data referring to single crystals showed that CBN has almost no random occupation in the pentagonal channels (A2 site), which are mainly filled by Ba [10], while only Ca atoms are present in the square channels (A1 site). However, analysis of the phonons in CBN-32 [11] showed that two-mode behaviour is present at frequencies below 150 cm$^{-1}$ (due to external vibrational modes from cations in the channels, i.e. out of the octahedra network), which points to double occupancy of the pentagonal site A2, with a partial occupation by Ca. X-ray experiments in powder also found a small amount of Ca in the A2 site, ~8% for CBN-31 [12].

It is interesting to note that the substitution of Sr by Ca does not lead to relaxor behaviour. CBN exhibits a dielectric behaviour similar to that found in the ferroelectric SBN compositions, although the ferroelectric phase transition shows some signs of diffusivity, such as a slightly broadened permittivity maximum [4]. This effect is probably attributable to the fact that CBN has smaller domains [13, 14] and shows a higher degree of structural disorder than SBN-35. However, a temperature dependent dispersion of the permittivity at high temperatures (~350 °C) for very low frequencies was found in CBN-28 [15] whose origin is unclear and led to the thinking that above $T_C$ the crystal was a relaxor, although the effect could be merely extrinsic, due to defects. Brillouin measurements found an intense broad central peak caused by polarization and strain fluctuations in the vicinity of $T_C$ [16]. Finally recent electron microscopy and diffraction studies [17] found that in the paraelectric phase there are incommensurate modulations that only disappear 200 degrees above $T_C$. All this experimental evidence points towards the presence of a proper paraelectric phase in CBN at much higher temperatures than $T_C$, as it was already found for ferroelectric SBN-35 [5].

The aim of this paper is to explain the dielectric behaviour of $(Ca_{0.3}Ba_{0.7})Nb_2O_6$ (CBN-30) over an ultrawide frequency range in relation to its phase transition, and to analyze the relaxations in comparison with the multiple polarization mechanisms as identified in the related SBN compound.

**EXPERIMENTAL DETAILS**

The CBN single crystals with x = 0.30 (CBN-30) were grown by the Czochralski method. Details about the growth can be found elsewhere [4, 18]. Samples with various geometries were cut from a big bulk crystal to fit the experimental requirements of each technique: time-domain terahertz transmission spectroscopy (TDTTS) from 0.2 to 2.5 THz, waveguide technique from 25 to 50 GHz, high-frequency coaxial line technique (1 MHz – 1.8 GHz), and low-frequency dielectric measurements (100 Hz – 1 MHz).

Far infrared (FIR) reflectivity spectra were acquired using a Fourier spectrometer Bruker IFS 113v with pyroelectric detectors as well as a He-cooled (1.5 K) Si bolometer. Room temperature spectra were measured in the range 30–3000 cm$^{-1}$ (~ $1\times10^{12}$ – $1\times10^{14}$ Hz). For low temperature measurements (from 300 to 10 K), a continuous-flow Oxford Optistat CF cryostat with the sample mounted in an He-gas bath was used. Due to the cryostat windows the upper frequency range was cut to 600 cm$^{-1}$. To polarize the light, we utilized a metal-mesh polarizer deposited on a thin polyethylene foil. For high temperatures (300–700 K), an adapted commercial high-temperature cell Specac P/N 5850 was placed in the spectrometer.

TDTTS measurements were carried on a thin polished plane parallel plate (dimensions 4×4×0.07 mm$^3$) with the polar axis parallel to the plate in the temperature range 10–510 K. We used a polarized electromagnetic field to measure the response along the polar *c*-axis from 0.2 to 2.5 THz with a resolution of 0.05 THz in a custom-made time-domain THz transmission spectrometer. An Optistat CF cryostat with Mylar windows was used for measurements down to 10 K. An adapted commercial high-temperature cell Specac P/N 5850 without windows was used to heat the sample.

Low-frequency dielectric measurements (1 Hz – 1 MHz) were performed with a Novocontrol Alpha-AN High-Performance Frequency Analyzer. Gold and silver electrodes were sputtered onto the faces of (001) oriented plates (4×4×0.5 mm$^3$), that were heated and then cooled with liquid nitrogen from 600 to 80 K in a THM-600 Linkam cell, at a temperature rate of 2.5 K/min under a probing field of 20 V/cm.

Dielectric properties in the high-frequency range (1 MHz – 1 GHz) were determined by measuring the complex reflection coefficient from the coaxial line loaded with the sample. The complex reflection coefficient was measured using an Agilent 8714ET vector network analyzer. The complex dielectric permittivity was calculated taking into account an inhomogeneous electric field distribution within the sample. The details of the so-called 'multi-mode' capacitor model used for the calculations can be found in the literature [19]. Data were acquired on a cooling cycle by keeping the cooling rate lower than 1 K/min. Samples were coated with silver electrodes. Several samples with different surface areas S were measured (S ≈ 3.34; 2; 0.54; 0.23 mm$^2$). The thickness of all the samples was 0.5 mm.

The 25–50 GHz frequency range was covered by two separate waveguide systems on needle-shaped samples. As the dielectric results vary significantly with temperature, several samples with different cross-section geometries were necessary to maximize the accuracy at different temperature intervals. The samples were hand-polished to the required dimensions using sand paper. The cross section of the samples varied from 0.07×0.01 – 0.12×0.12 mm$^2$ and they were placed in the middle of the wider wall of the waveguide with the *c*-axis perpendicular to it (electric field parallel to the *c*-axis of the sample). The scalar reflection and transmission coefficients were measured by an Elmika R2400 scalar network analyzer. More details can be found in [8, 20].

Raman spectra were measured in the backscattering geometry using a Renishaw RM-1000 Raman microscope, equipped with an Ar laser (514.5 nm wavelength, ~25 mW power) and two Bragg filters, allowing the spectra to be recorded from 2 cm$^{-1}$. The diameter of the laser spot on the sample surface was ~10 μm through a 20× objective and the spectral resolution was about 1.3 m$^{-1}$ with a 2400 l/mm grating. Spectra were collected from 700 K down to 80 K in a THM-600 Linkam cell using liquid nitrogen as a coolant.

**EXPERIMENTAL RESULTS**

The structure of CBN is depicted in Fig. 1 a,b, at room temperature, using X-ray diffraction data on single crystals [10]. TTBs are known for having two different types of oxygen octahedra according to the different atomic positions of the cation located inside. The so-called linking octahedra, in dark color, have Nb(1) atoms at 2*b* sites, and the perovskite-like unit octahedra have Nb(2) atoms at 8*d* sites. The figure shows how the linking octahedra (dark color) form independent columns along the *c*-axis, whilst the others form a perovskite-like unit of interconnected octahedra (light color). In addition, linking octahedra show almost no deformation (no oxygen tilts).

Various experiments were carried out to investigate the complex dielectric behaviour of CBN, and the results are shown below.

Dielectric data

The dielectric response at selected frequencies from ~1 Hz to 28 GHz is displayed in Fig 1c. Experimental data were obtained on cooling along the polarization axis. The temperature dependence of the permittivity ε' and the dielectric loss ε'' at selected frequencies are both shown in Fig 1c: lines for low frequencies (1 Hz – 1 MHz); symbols for high frequencies (1 MHz – 2.8 GHz). Both sets of data agree well within experimental uncertainties. The experiment at 28 GHz was done using the waveguide technique in a very small needle, therefore only data below 470 K were sufficiently reliable.

The CBN-30 crystal shows a clear anomaly at $T_C$ ~ 495 K. The maxima of the permittivity are not completely sharp, which speaks in favor of a phase transition of prevailing second order with some diffuse character. No further anomaly is seen down to low temperatures. The small shoulder in ε' near 400 K at frequencies below 10Hz is due to a thermally activated process of extrinsic origin and not related with the lattice dynamics of the crystal. The very weak shoulders seen in the dielectric loss ε'' below room temperature refer to very slow dissipation processes present in the sample without contributing to the permittivity or to the polarization of the crystal. Above $T_C$ there is no evidence of relaxor behaviour, as the maxima of ε' do not shift up with increasing frequency. In the dielectric loss, the effect of the conductivity at high temperatures is seen gradually below ~ 10 kHz masking the phase transition anomaly. The increase of permittivity at ~1 Hz towards 600 K reminds the effect found in CBN-28 [15], however at such high temperatures extrinsic effects, due to oxygen vacancies for instance, can be responsible for it, and this cannot be taken as a proof of relaxor behaviour.

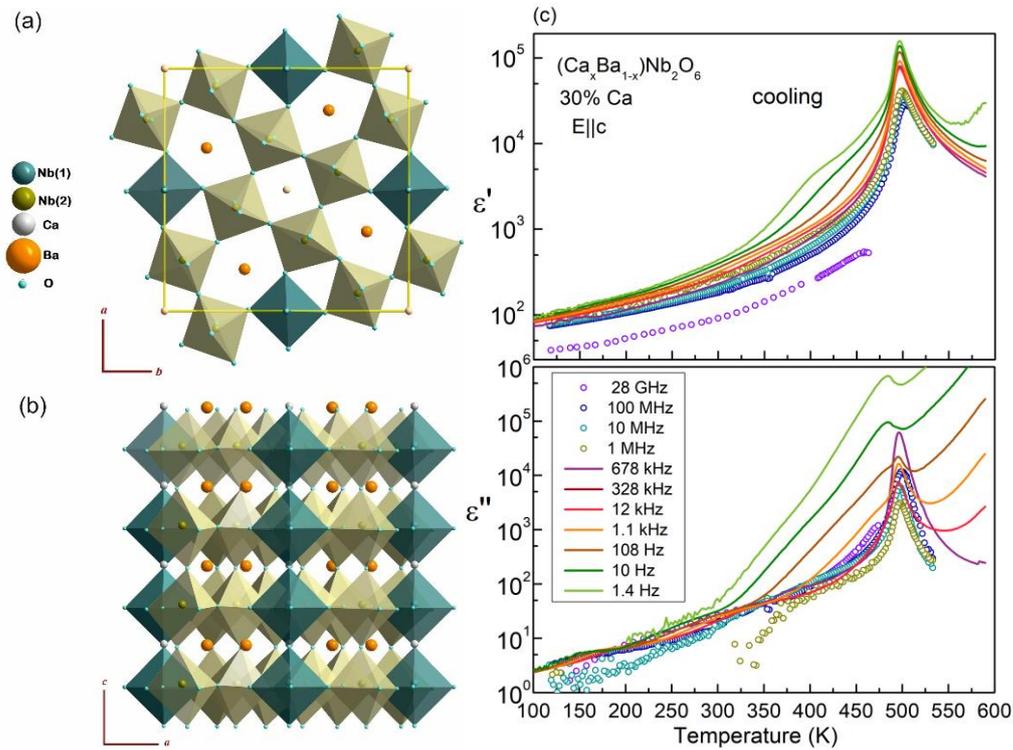

Figure 1: Structure of CBN-30 in two orientations (a,b) and dielectric data (c) at several frequencies (note the logarithmic scales in ε' and ε''). Nb(1) are inside the linking octahedra in dark color, and Nb(2) inside the perovskite-like octahedra in light color. Low frequencies (1 Hz – 1 MHz) – lines; high frequencies (1 MHz – 2.8 GHz) – symbols. The uncertainty bars are smaller than the size of graphic points.

Infrared spectra

According to the published results of the site group analysis of the lattice vibrations for TTBs [5, 11], for CBN there are 46 atoms in the unit cell (in the case of single atom occupation in pentagonal A2 sites) and the polar vibrations active along the *c*-axis are 9 $A_{2u}$(c;–) in the paraelectric phase, and 19 $A_1$(c; $a^2+b^2,c^2$) in the ferroelectric phase. Infrared (IR) and Raman activities are given in brackets. However, if we consider the two-atom occupancy in the pentagonal channels, with Ca and Ba, the number of vibrations coming from these two atoms are doubled. In this case, there are effectively 50 atoms in the unit cell and we expect 10 $A_{2u}$(c;–) modes in the paraelectric phase and 21 $A_1$(c; $a^2+b^2,c^2$) modes in the ferroelectric phase. Out of them, one $A_{2u}$ and one $A_1$ are acoustic phonons, three $A_{2u}$ modes and five $A_1$ modes belong to the external vibrations –due to Ba and Sr vibrations– corresponding to phonons with frequencies below ~180 cm$^{-1}$, and the rest are modes due to Nb–O stretching and bending vibrations in the octahedral network.

The FIR reflectivity of the CBN-30 crystal was measured from 600 K to 10 K and the obtained spectra along the polar axis, showing $A_{2u}/A_1$ symmetry modes, are shown in Fig. 2a together with the data collected by the TDTTS experiment. The reflectivity spectra are normalized with respect to the points measured in the THz range and the corresponding fits at selected temperatures are shown. Reflectivity is calculated by $R(\omega) = \left|\frac{\sqrt{\hat{\varepsilon}(\omega)}-1}{\sqrt{\hat{\varepsilon}(\omega)}+1}\right|^2$, being $\hat{\varepsilon}(\omega)$ the complex dielectric function. Due to the presence of very broad and asymmetric bands, the spectra were fitted using the generalized or 4-parameter oscillator model with the factorized form of the dielectric function [21]:

$$\hat{\varepsilon}(\omega) = \varepsilon'(\omega) - i\varepsilon''(\omega) = \varepsilon_\infty \prod_{i=1}^{n} \frac{\omega_{LOi}^2 - \omega^2 + i\omega\gamma_{LOi}}{\omega_{TOi}^2 - \omega^2 + i\omega\gamma_{TOi}} \quad (1a)$$

$$\Delta\varepsilon_i = \frac{\varepsilon_\infty}{\omega_{TOi}^2} \frac{\prod_j (\omega_{LOj}^2 - \omega_{TOi}^2)}{\prod_{j \neq i} (\omega_{TOj}^2 - \omega_{TOi}^2)} \quad (1b)$$

where $\varepsilon_\infty$ is the permittivity at frequencies much higher than the phonon frequencies, $\omega_{TOi}$ and $\omega_{LOi}$ are the transverse and longitudinal frequencies of the *i*-th phonon mode, and $\gamma_{TOi}$ and $\gamma_{LOi}$ their respective damping parameters. The real and imaginary part of the dielectric permittivity calculated from these fits are shown together with the experimental TDTTS points in Fig. 2 b, at selected temperatures.

Below the phonons, there is an additional mode to fit the THz data, and in addition, at lower frequencies another contribution should be included to obtain approximate values of the permittivity in the MHz–GHz range.

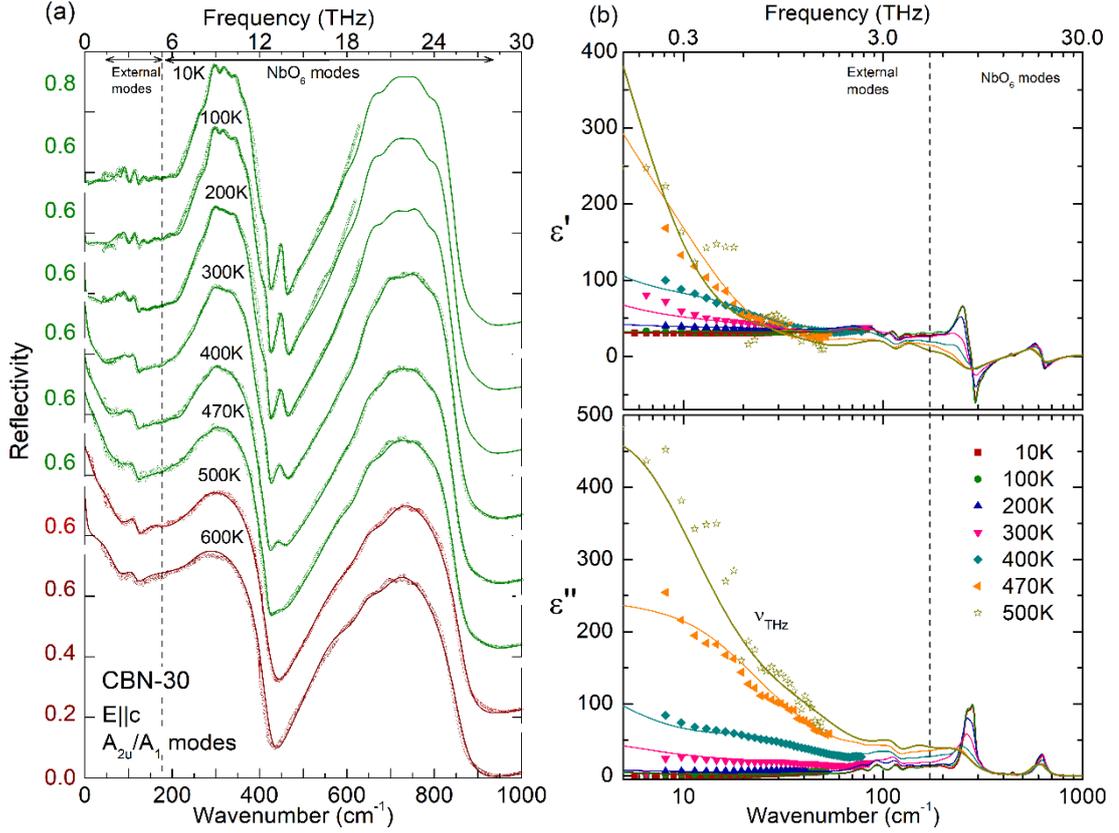

Figure 2: (a) FIR spectra (symbols) at different temperatures with their fits (lines) using eqs.(1) corresponding to the $A_{2u}/A_1$ polar modes. (b) Permittivity and dielectric loss obtained from the fit (lines) together with experimental data points from the TDTTS experiment (symbols). Note the logarithmic scale in panel b. The uncertainty bars are smaller than the size of graphic points.

The phonon contribution to the permittivity along the polar axis is rather low ($\varepsilon_c' \sim 40$ above room temperature), as in other TTB materials [6, 22], but the anharmonic mode in the THz range, called central mode, dramaticalFly enhances the permittivity $\varepsilon'$ values in the THz range, close to 10 cm$^{-1}$, at high temperatures (see Fig. 2b). On cooling, the contribution of the central mode to the permittivity almost vanishes, so that the permittivity at THz frequencies practically coincides with the phonon contribution. In Fig. 2b the dielectric loss spectra at different temperatures are also plotted in the lower panel. Each peak or maximum corresponds to a phonon mode. On cooling, all the phonons are better distinguished and, in the THz range and at high temperatures, the intense central mode (labelled $\nu_{THz}$) is seen as a broad feature, which gradually weakens below $T_C$ and shifts to higher frequencies. The remaining phonons do not show any significant softening. Their frequencies and dielectric contribution (separated in external modes and internal vibrations of the NbO$_6$ octahedra) are presented in Table I.

Raman spectroscopy

The polarized Raman spectra were measured on the same plate as the one used for the IR experiment. A$_1$ transverse optic (TO) modes were obtained on the plate with the polarization in plane in X(ZZ)X̄ geometry (backscattering). As these modes are simultaneously IR and Raman active their parameters can be directly compared. This is particularly useful since the IR phonons above 600 cm$^{-1}$ cannot be measured below room temperature due to the opacity of the cryostat windows. By combining both techniques all the polar A$_1$ modes could be identified.

The temperature dependence of the spectra is shown in Fig. 3a corrected for the Bose-Einstein temperature factor and artificially shifted. Because the Raman scattering cross section is proportional to the imaginary part of the response function, the fluctuation dissipation theorem allows us to compare Raman and IR data using the formula [23]

$$I_s(\omega) = \hbar/\pi \, [n(\omega)+1] \chi''(\omega),$$

where $I_s$ is the Stokes Raman scattered intensity, $n(\omega)$ the Bose–Einstein thermal factor, $\hbar$ is the reduced Planck constant, and $\chi''(\omega)$ is the imaginary part of the susceptibility, which in our case is directly related to the complex dielectric function by $\hat{\varepsilon}(\omega) = \varepsilon_0(1 + \hat{\chi}(\omega))$.

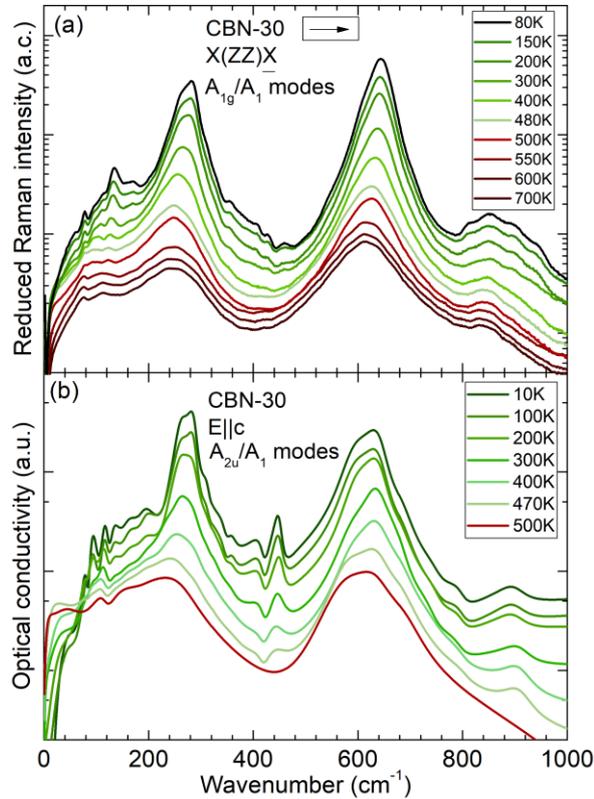

Figure 3: Raman spectra of $A_1$(TO) modes (a), corrected for the Bose-Einstein thermal population factor and optical conductivity (b), calculated from the fits to the IR reflectivity. Note the logarithmic scales used to see well the weak modes.

Figure 3b shows the optical conductivity of the IR modes obtained from the fit (Eq.(1)). This quantity is proportional to the dielectric loss, $\sigma'(\omega) = 2\pi c \varepsilon_0 \omega \varepsilon''(\omega)$, and it shows better the higher frequency IR modes. $A_1$ modes obtained by both techniques reveal a good agreement. The main discrepancy is seen in the oscillator strength of some of the weak modes, which differs in each technique. However, the phonon frequencies should not differ much, as $A_1$ modes are both IR and Raman active.

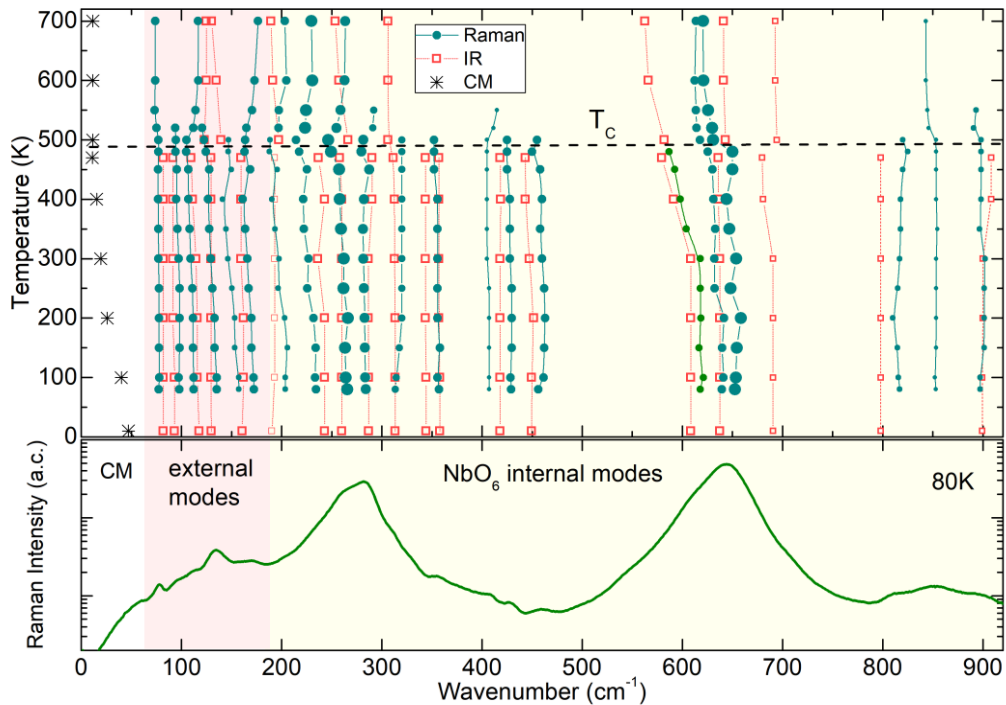

Figure 4 Temperature dependence of the frequencies of the polar phonons ($A_{2u}/A_1$ IR + $A_{1g}/A_1$ Raman) and the central mode, together with the B-E reduced Raman spectra at 80 K in logarithmic scale to see the very weak modes. Uncertainty in frequencies lies within the size of graphic points.

The frequencies obtained by Raman and FIR techniques are plotted in Fig. 4, showing a very good agreement in the ferroelectric phase, where the $A_1$ modes are simultaneously IR and Raman active. In the paraelectric phase, IR and Raman active modes belong to different representations ($A_{2u}$-IR and $A_{1g}$-Raman) and they cannot be correlated. The presence of several weak polar modes at ~ 550 K is attributed to the presence of polar fluctuations in the paraelectric phase when approaching the ferroelectric phase transition, as a precursor effect. In the ferroelectric phase the number of external vibrations –due to the cations in the channels and present below about 180 cm$^{-1}$– is five instead of three, as expected from factor group analysis. This indicates that the pentagonal site is occupied partially by Ca, not only Ba, as pointed out for CBN-32 [11]. An extra Raman mode close to 150 cm$^{-1}$ is probably due to the presence of the oxygen octahedra tilt, found in similar compounds with Nb octahedra networks [11, 24]. The rest of the modes correspond to internal vibrations of the oxygen octahedra, and they are common to other Nb based TTBs and layered perovskite $Sr_2Nb_2O_7$ [25].

In Table I we present the IR phonon parameters for the E||c polarization at 500 K in the paraelectric phase ($A_{2u}$ symmetry) and at 100 K in the ferroelectric phase ($A_1$ symmetry). The central mode $\nu_{THz}$ is added to the table, but as it is overdamped, we show its renormalized frequency $\nu_{THz}^2/\gamma_{THz}$, which corresponds to its maximum in the dielectric loss. For a more quantitative comparison, we add the frequencies of the Raman modes in the ferroelectric phase too.

The external modes from cationic vibration appear below 180 cm$^{-1}$. Their overall dielectric strength is 14, which is higher than the contribution from all the internal modes ($\Delta\varepsilon$ = 12). The main phonons (with higher dielectric strength and Raman area) are shown in bold, and they correspond to the main Nb-O stretching vibrations, present in the range 200–280 cm$^{-1}$. There are many other weaker modes detected complementary by both techniques. In general, all frequencies found by both techniques match reasonably. The exceptions are a weak mode detected by IR spectroscopy at 343 cm$^{-1}$, which is absent in Raman, probably due to its even weaker Raman signal, and two Raman vibrations (also weak) with frequencies ~407 and ~851 cm$^{-1}$, not visible in the IR spectra. The higher frequency one could be related with the breathing of the oxygen octahedra in the perovskite part of the network, which is known to be non-polar and therefore IR inactive [26].

**DISCUSSION**

To understand the overall dielectric behavior of CBN-30, the information of the excitations present in the structure below the phonon frequencies is crucial, as they carry the main contribution to the permittivity. The frequency dependence of the complex dielectric function $\hat{\varepsilon}(\nu)$ was studied at different temperatures in a broad frequency range from Hz to THz. This frequency dependence was fitted using a model based on a sum of Cole-Cole relaxations, an overdamped oscillator in the THz range and the contribution of phonons and electrons:

$$\hat{\varepsilon}(\nu) = \varepsilon'(\nu) - i\varepsilon''(\nu) = \sum_j \frac{\Delta\varepsilon_{0j}}{1+(i\nu/\nu_{0j})^{1-\alpha_j}} + \frac{\Delta\varepsilon_{THz}\nu_{THz}^2}{\nu_{THz}^2-\nu^2+i\gamma_{THz}\nu} + \sum_{ph}\Delta\varepsilon_{ph} + \varepsilon_\infty \qquad (2)$$

In eq. (2), $\Delta\varepsilon_{0j}$ is the dielectric contribution of the relaxation (*i.e.* relaxation step), $\nu_{0j}$ its characteristic relaxation frequency, and $\alpha_j$ a real index between 0 and 1, accounting for the deviation from the pure Debye character ($\alpha = 0$). In the oscillator term, $\Delta\varepsilon_{THz}$ is the contribution of the central mode, $\nu_{THz}$ is its frequency and $\gamma_{THz}$ refers to its damping coefficient. A sum of Lorentzians $\Sigma\Delta\varepsilon_{ph}$ was used to account for the contribution of phonons and $\varepsilon_\infty$ for the electronic contribution.

In Fig. 5 the frequency dependence of the combined dielectric data, from 100 Hz to 1 THz is depicted together with the fit obtained by eq.(2). Experimental data are the same as those in Figs. 1c and 2b. The merged dielectric data reveal a main excitation at high temperature in the paraelectric phase below 550 K, in the $10^8$ Hz region, labelled $\nu_{01}$. Below 1 MHz, there is another weaker excitation, $\nu_{02}$, and above 1 THz, the central mode feature (labelled $\nu_{THz}$) is present.
On cooling the excitation $\nu_{01}$ broadens, weakens and splits into several contributions: $\nu_{01a}$, $\nu_{01b}$ and $\nu_{GHz}$. The fit of the dielectric response shows that each of these contributions slows down towards lower frequencies, except for the excitation present in the GHz range, which remains in this frequency range. This behaviour indicates that multiple mechanisms coexist and account simultaneously for the ferroelectric behaviour in CBN-30.

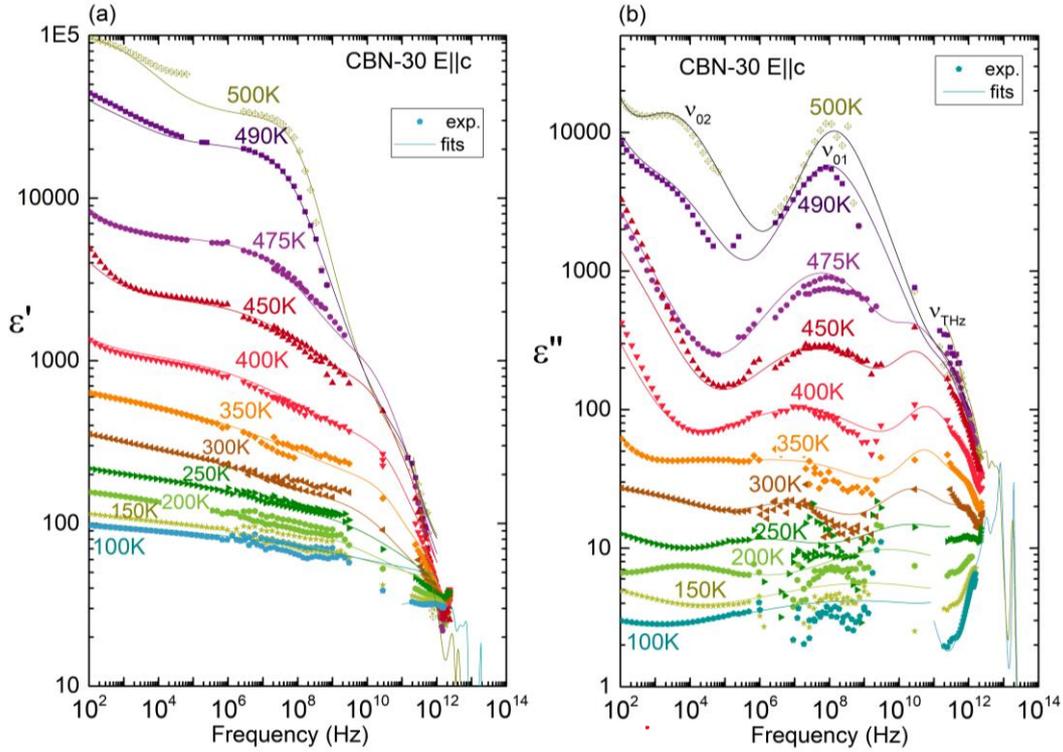

Figure 5 Frequency dependences of the permittivity $\varepsilon'$ (a) and loss $\varepsilon''$ (b) of CBN-30 along the polar axis at different temperatures together with their fits. Experimental data – open symbols, fits with eq.(2) – lines. Note that the spectra are not artificially shifted. The additional fits at 500 K and 100 K above $10^{11}$ Hz correspond to the fits of the phonons using eq.(1). Experimental error lies within the size of graphic points.

The temperature dependence of the parameters of the corresponding excitations: frequencies $\nu$ and dielectric contributions $\Delta\varepsilon$ are presented in Fig. 6 (and in Table I for comparison with phonons). Several excitations with different temperature responses are found. The temperature dependence of their frequencies (Fig. 6a) shows different behavior for $\nu_{THz}$, $\nu_{01}$ (the main contribution) and $\nu_{02}$. Below $T_C$, $\nu_{01}$ splits into $\nu_{GHz}$, $\nu_{01a}$ and $\nu_{01b}$. The discrepancy in the temperature evolution of all the frequencies indicates that each of the excitations has a different microscopic origin and different correlation length and it reflects the different kinetics in the development of the polar domains or nanodomains. In Fig. 6b the contribution to the permittivity of each excitation is plotted along with the measured permittivity at several frequencies. This plot illustrates that the weakest contribution comes from phonons, while the main contribution comes from the $\nu_{01a}$ excitation which slows down towards the kHz range. This relaxation is the main polarization mechanism, as in other TTBs. Its microscopic origin cannot be assigned from our dielectric measurements, however from the investigations in SBN, it can be related to Nb dynamics.

The presence of acoustic modes in the GHz range [14, 16, 27] and the small size of the domains and domain walls found by PFM for CBN [13, 14] support the attribution of $\nu_{GHz}$ to domain wall oscillations, as in other ferroelectric materials [28, 29]; however, other mechanisms cannot be excluded.

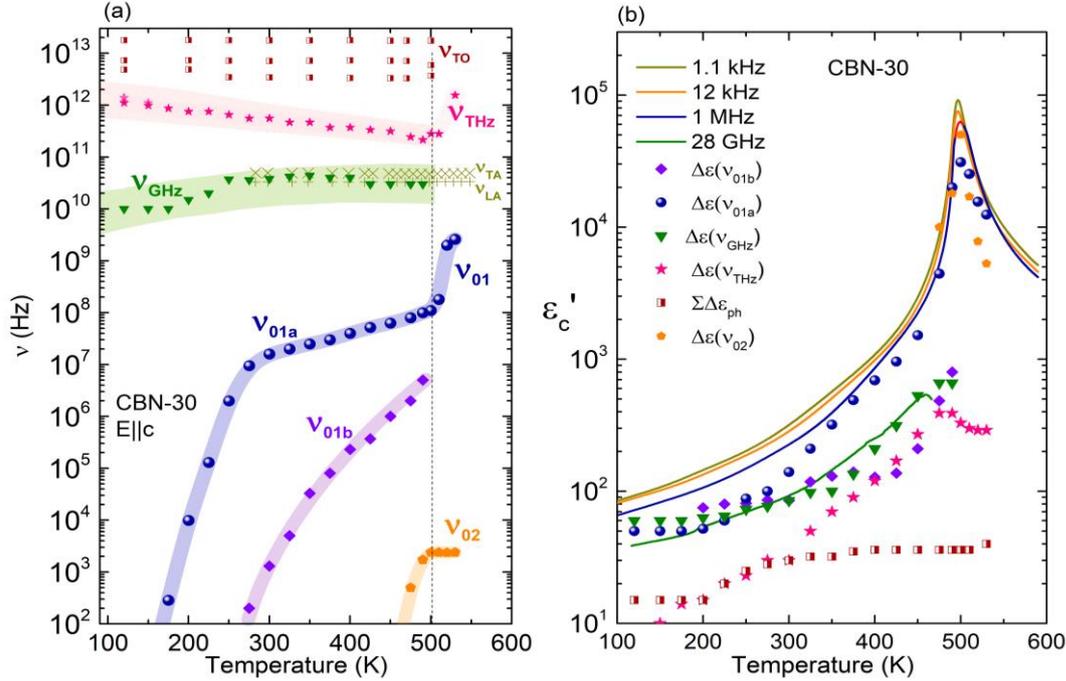

Figure 6 Temperature dependence of the frequencies of the main excitations in CBN-30 (a), in logarithmic scale, and their dielectric contribution to the permittivity (b), together with selected experimental data. $\nu_{TA}$ and $\nu_{LA}$ taken from [27]. Uncertainty lies within the colored zones or the size of graphic points.

The behavior of the permittivity and the spontaneous polarization in TTB family [2] seems to indicate that the ferroelectric phase transition is of second order. However, there are also some indications of weak first order behaviour. For instance in the volume of the unit cell [27,30] and in the small thermal hysteresis found by calorimetry [14]. Therefore the ferroelectric phase transition in CBN shows a mixed behaviour: second order with a weak first-order component, and also a mixed mechanism between displacive (soft central mode) and order-disorder (relaxation slowing down). Below $T_C$ the relaxation splits and each of the new relaxations presents different temperature behaviour, which is a peculiarity of the TTB structure.

Comparison to SBN

The dielectric behaviour of CBN resembles well that found in ferroelectric SBN [8], the main difference being the higher transition temperature of CBN, and the frequency of some excitations. Compared to the ferroelectric SBN compositions, $\nu_{THz}$ and $\nu_{GHz}$ display similar temperature behaviour and frequency, mimicking the behaviour in SBN. The temperature dependence of $\nu_{THz}$ was fitted well to the critical law $\nu_{THz} = a(T_1 - T)$, which correspond to the renormalized Cochran law, as in SBN compositions [8]. The values found were $T_1 = 548 \pm 19$ K and $a = (2.3 \pm 0.2) \times 10^9$ s$^{-1}$K$^{-1}$, similar to those found for SBN. This soft central mode also corresponds well with the frequencies of the central peak found in Brillouin scattering in CBN-28 [16], however, due to the different techniques employed, the microscopic origin of both anharmonic features could be different.

The parameters of the main excitations in both systems are plotted in Fig. 7. However, as SBN has lower $T_C$, for graphical comparison, $T_C$ has been artificially shifted by 40, 110 and 130 degrees in SBN-35, -51 and -61, respectively. In this way, it is easier to identify the behaviour of each excitation with respect to the phase transition rather than the actual temperature. The behaviour of $\nu_{THz}$, $\nu_{GHz}$ and $\nu_{01a}$ in CBN-30 resembles mostly that of SBN around the phase transition (see Fig. 7). There are, nevertheless, some interesting differences.

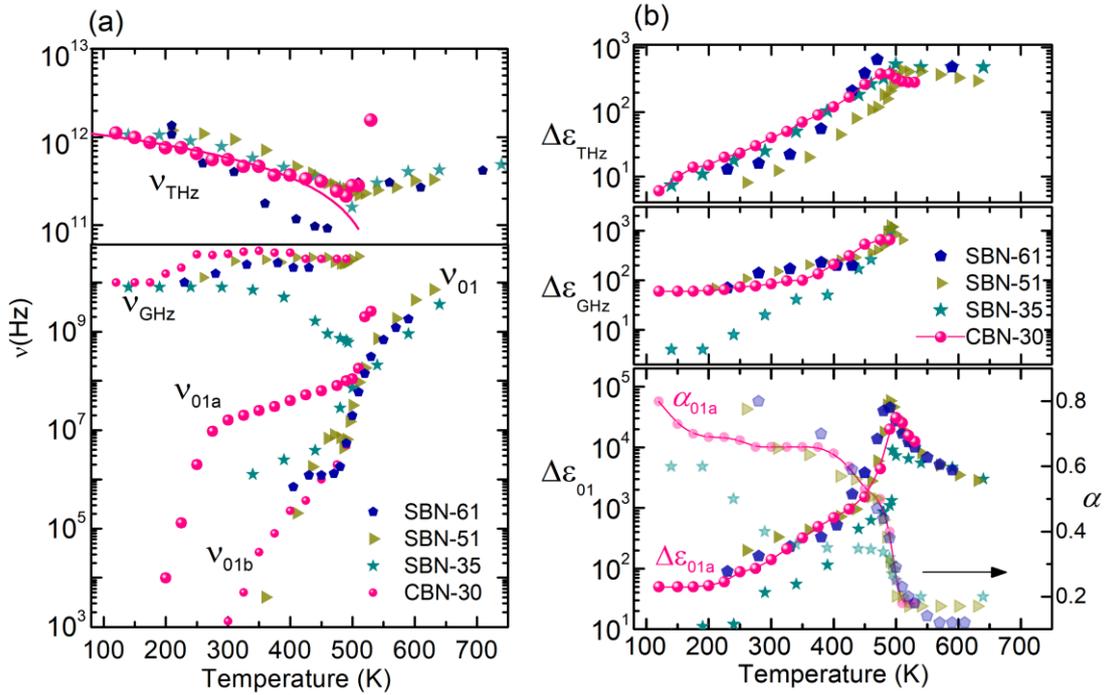

Figure 7: Qualitative comparison of the parameters of the main excitations found in several TTBs. The line corresponds to the fit of $\nu_{THz}$ using the equation $\nu_{THz}=a(T_1-T)$. Note that the real transition temperatures of the three SBN crystals were artificially

shifted for comparison purposes. For the relaxation $\nu_{01a}$, the parameter $\alpha$ was added to $\Delta\varepsilon_{01a}$. Experimental error lies within the size of graphic points.

For instance, the frequency and dielectric strength of $\nu_{GHz}$ agree with those of SBN, except for the composition SBN-35, where both parameters are lower. This is due to the presence of larger domains and thicker domain walls in SBN-35. The other crystals: CBN-30, SBN-51 and SBN-61 have much smaller domains [13,14].

The most obvious difference with SBN is the frequency of the main excitation, the relaxation $\nu_{01}$, which splits into three components $\nu_{GHz}$, $\nu_{01a}$ and $\nu_{01b}$ below the phase transition. This effect suggests the coexistence of domains of different sizes, with the main drivers of the polarization response being those with higher relaxation frequency $\nu_{01a}$ and, probably, smaller size. The secondary component $\nu_{01b}$ is weaker (as seen in Fig. 6b), so the main relaxation $\nu_{01a}$ has similar dielectric strength as in SBN (see Fig. 7b), but with an order of magnitude higher frequency (Fig. 7a). This may be related to the microstructure of the polar domains in CBN, which have a granular texture with a wide range of sizes [14] and remains to be investigated.

As in other TTB structures, in CBN, there are chains of two types of octahedra contributing to the dielectric response. The linking octahedra $Nb(1)O_6$ (darker color in Fig. 1a) form independent columns along the *c*-axis. They show almost no distortion (no oxygen tilts) and should represent the fixed part of the polarization (independent on temperature), as found in SBN [8]. On the contrary, the perovskite-like octahedra $Nb(2)O_6$ (in light color in Fig. 1a) depict the octahedra with dynamic polarization, which is affected by the temperature through the Nb(2) shifts. At room temperature, the shifts of both Nb(1) and Nb(2) atoms along the polar direction in CBN-31 are comparable: 0.174 Å for Nb(1) and 0.146 Å for Nb(2) [12], as in the ferroelectric SBN compositions [8]. This may be the reason why CBN does not show relaxor behaviour despite the high level of atomic disorder found, particularly in the oxygen atoms [12], deserving further investigation. Analysis of diffuse scattering in CBN should help to reveal the presence of different types of local ordering.

Our experimental results show that the phase transition in CBN is ferroelectric without relaxor character and is caused by several coexisting mechanisms.

At the atomic level, there is a tendency of the Nb atom to shift along the polar axis, but this mechanism is weak and does not contribute significantly to the permittivity to produce a classical soft phonon mode behavior. On the contrary, anharmonic cationic effects are stronger and they are revealed by an excitation in the THz range, the central mode $\nu_{THz}$, which already shows an important anomaly at $T_C$. This central mode is probably caused by the dynamic disorder of the cations, supported by the high anisotropic displacements found in the diffraction experiment.

Nevertheless, the main actor during the phase transition is a relaxation located in the GHz range in the paraelectric phase, $\nu_{01}$, which has the strongest contribution to the permittivity – $\Delta\varepsilon(\nu_{01}) \sim 10000$ just above $T_C$– and it is related to polar fluctuations caused by Nb atoms dynamics. This excitation slows down on cooling and splits into several excitations below $T_C$:

-A relaxation with frequency close to 10 GHz ($\nu_{GHz}$), consistent with the oscillations of the ferroelectric domain walls.

-A relaxation in the sub-GHz range ($\nu_{01a}$), which slows down to kHz at low temperatures, and shows the strongest dielectric anomaly, being the main cause of the permittivity. Due to the large Nb shifts it appears that the Nb-O chains along the polar axis have a relatively large correlation length and their dynamics should have an important influence on this excitation and on the dynamics of some of the ferroelectric domains.

-A weaker relaxation in the sub-MHz range ($\nu_{01b}$), which also slows down to at least 10 Hz, which is associated with the dynamics of larger ferroelectric domains.

In addition, another relaxation $\nu_{02}$ is present at lower frequencies, although it is only seen around $T_C$, before shifting to frequencies below 100 Hz. This relaxation could be due to domain breathing, as found in SBN-81 [31].

**CONCLUSIONS**

The ferroelectric phase transition of the tetragonal tungsten bronze CBN with 30 % Ca shows a complex behavior involving several excitations. Various polarization mechanisms with different correlation lengths and characteristic frequencies are found, as in the related material SBN. These mechanisms coexist and contribute simultaneously to the overall dielectric response.

The absence of a classical soft phonon mode is a feature of the TTB structure, due to structural constraints that prevent long-range correlation of the Nb atoms. However, the material exhibits a strong soft anharmonic feature, the central mode, related to the anharmonic motion of the cations.

No traces of relaxor behaviour were found in the paraelectric phase below 600 K, but a strong relaxation in the GHz range with a maximum contribution to the permittivity, probably due to polarization fluctuations of nanometric size, gradually weakens and finally splits below $T_\mathrm{C}$. Four excitations related to different polarization correlation lengths appear in the ferroelectric phase. The main one, slowing down from the sub-GHz to the kHz range, is attributed to the dynamics of the polar domains due to the Nb-O chains along the polar axis, as in SBN. Other weaker contributions in the GHz range and below MHz can be attributed to domain wall oscillations and dynamics of larger size ferroelectric domains.

**Acknowledgment**

This work was partially supported by the Czech and Lithuanian Academies of Sciences through the project LAS-21-02.

E.B. acknowledges support from the Ministry of Education, Youth, and Sports of the Czech Republic by the EU co-funded grant "Ferroic Multifunctionalities", Project No. CZ.02.01.01/00/22_008/0004591.

**Table I**

Parameters of the IR phonons of CBN-30 for the E||c polarization in the paraelectric and ferroelectric phases ($\nu_T$ and $\gamma_T$ in cm$^{-1}$, their uncertainty is ~ 0.5 cm$^{-1}$). For 100 K the Raman frequencies are also added. The central mode $\nu_{THz}$ is shown by its renormalized frequency $\nu_{THz}^2/\gamma_{THz}$ with an asterisk. The dielectric strength $\Delta\varepsilon$ is unitless and its values have an uncertainty of 0.5. The most intense modes are in bold.

|  | **CBN-30**   *E*\|\|*c* | | | | | | |
|---|---|---|---|---|---|---|---|
|  | Ferroelectric phase (100 K) $A_1$ modes | | | | Paraelectric phase (500 K) $A_{2u}$ modes | | |
|  | $\nu_{Ram}$ | $\nu_{TOi}$ | $\gamma_{TOi}$ | $\Delta\varepsilon_i$ | $\nu_{TOi}$ | $\gamma_{TOi}$ | $\Delta\varepsilon_i$ |
| $\nu_{THz}$ |  | 46.7* | -- | 4 | 11.5* |  | 275 |
| External vibrations in channels | 77.3 | 83.1 | 14.1 | 0.8 |  |  |  |
|  | 97.7 | 91.9 | 19.4 | 4.2 |  |  |  |
|  | 112.1 | 118.1 | 19.4 | 2.9 | 123.1 | 43.2 | 8.4 |
|  | 135.2 | 132.8 | 27.9 | 2.7 | 139.4 | 112.5 | 37.4 |
|  | 157.2 |  |  |  |  |  |  |
|  | 172.4 | 162.3 | 44.4 | 3.3 |  |  |  |
| Internal vibrations of NbO$_6$ | 203.6 | 192.5 | 38.5 | 0.7 | 197.0 | 64.6 | 3.9 |
|  | 234.4 | 243.0 | 42.8 | 1.9 |  |  |  |
|  | **265.6** | **260.2** | **28.4** | **3.6** | 266.2 | 81.6 | 3.3 |
|  | **284.2** | **286.7** | **23.5** | **2.3** |  |  |  |
|  | 313.4 | 310.4 | 28.7 | 0.7 | 306.2 | 44.0 | 0.1 |
|  |  | 343.0 | 27.4 | 0.1 |  |  |  |
|  | 357.1 | 356.3 | 25.6 | 0.1 |  |  |  |
|  | 407.2 |  |  |  |  |  |  |
|  | 429.2 | 418.3 | 24.47 | 0.03 |  |  |  |
|  | 462.0 | 449.2 | 21.95 | 0.1 |  |  |  |
|  | 634.6 | 608.4 | 66.8 | 1.6 | 581.9 | 110.0 | 2.3 |
|  | 646.7 | 637.5 | 43.6 | 1.1 | 643.0 | 66.0 | 0.1 |
|  | 712.7 | 690.5 | 82.4 | 0.04 | 694.5 | 110.0 | 0.6 |
|  | 814.2 | 798.0 | 18.9 | 0.001 |  |  |  |
|  | 851.4 |  |  |  |  |  |  |
|  | 887.3 | 899.5 | 36.1 | 0.01 |  |  |  |

# References

1   M.D. Ewbank, R.R. Neurgaonkar, W.K. Cory and J. Feinberg, Photorefractive properties of strontium-barium niobate, *J. Appl. Phys.* **62**, 374 (1987).

2   A. M. Glass, Investigation of the Electrical Properties of $Sr_{1-x}Ba_xNb_2O_6$ with Special Reference to Pyroelectric Detection, *J. Appl. Phys.* **40**, 4699 (1969).

3   T. Łukasiewicz, M.A. Swirkowicz, J. Dec, W. Hofman, W. Szyrski, Strontium–barium niobate single crystals, growth and ferroelectric properties, *J. Crystal Growth* **310**, 1464 (2008).

4   O. V. Malyshkina, V. S. Lisitsin, J. Dec and T. Łukasiewicz, Pyroelectric and Dielectric Properties of Calcium Barium Niobate Single Crystals, *Phys. Solid State* **56**, 1824, (2014).

5   E. Buixaderas, M. Kempa, V. Bovtun, C. Kadlec, M. Savinov, F. Borodavka, P. Vaněk, G. Steciuk, L. Palatinus and J. Dec, Multiple polarization mechanisms across the ferroelectric phase transition of the tetragonal tungsten bronze SBN-35, *Phys. Rev. Mat* **2**, 124402 (2018).

6   E. Buixaderas, C. Kadlec, M. Kempa, V. Bovtun, M. Savinov, P. Bednyakov, J. Hlinka and J. Dec, Fast polarization mechanism in the uniaxial tungsten-bronze relaxor SBN-81, *Sci. Rep.* **7,** 18034 (2017).

7   M. Paściak, P. Ondrejkovic, J. Kulda, P. Vanek, J. Drahokoupil, G. Steciuk, L. Palatinus, T. R. Welberry, H. E. Fischer, J. Hlinka and E. Buixaderas, Local structure of relaxor ferroelectric $Sr_xBa_{1-x}Nb_2O_6$ from a pair distribution function analysis., *Phys. Rev. B* **99**, 104102 (2019).

8   E. Buixaderas, M. Kempa. Š. Svirskas, C. Kadlec, V. Bovtun, M. Savinov, M. Paściak and Jan Dec, Dynamics of mesoscopic polarization in uniaxial tetragonal tungsten-bronze $(Sr_xBa_{1-x})Nb_2O_6$, *Phys. Rev. B* , (2019).

9   V. Krayzman, A. Bosak, H.Y. Playford, B. Ravel and I. Levin, Incommensurate Modulation and Competing Ferroelectric/Antiferroelectric Modes in Tetragonal Tungsten Bronzes, *Chem. of Mat.* **34** (22), 9989 (2022).

10  M. Eßer, M. Burianek, P. Held *et al.*, Optical characterization and crystal structure of the novel bronze type $Ca_xBa_{1-x}Nb_2O_6$, *Cryst. Res. Technol.* **38**, 457 (2003).

11  E. Buixaderas and J. Dec, OAJ materials and Devices, vol 5(1) – chap 8 in "Perovskites and other Framework Structure Crystalline Materials", p 281 (Coll. Acad. 2021).

12  H.A. Graetsch, C.S. Pandey, J. Schreuer, M. Burianek and M. Muhlberg, Incommensurate modulations of relaxor ferroelectric CBN24 and CBN31, *Acta Cryst. B* **70**, 743 (2014).

13  V.V. Shvartsman, D. Gobeljic, J. Dec and D.C. Lupascu, A Piezoresponse Force Microscopy Study of $Ca_xBa_{1-x}Nb_2O_6$ Single Crystals, *Materials* **10**, 1032 (2017).

14  E. Buixaderas, P. Bérešová, P. Ondrejkovič, P. Vaněk, M. Savinov, P. Bednyakov, J. Dec, D. Mareš, M. Ševčík and M. Landa, Acoustic phonons in unfilled tetragonal tungsten-bronze crystals, *Phase Transitions* **91**, 976 (2018).

15  Y.J. Qi, C.J. Lu, J. Zhu, X.B. Chen, H.L. Song, H.J. Zhang, X.G. Xu, Ferroelectric and dielectric properties of CBN-28 single crystals of tungsten bronzes structures, *Appl. Phys. Lett.* **87**, 082907 (2005).

16  K. Suzuki, K. Matsumoto, J. Dec, T. Łukasiewicz, W. Kleemann, and S. Kojima, Critical slowing down and elastic anomaly of uniaxial ferroelectric $Ca_{0.28}Ba_{0.72}Nb_2O_6$ crystals with tungsten bronze structure, *Phys. Rev. B* **90**, 064110 (2014).

17  R. Beanland, L. Harrison, S. Khan, T, Brown, T. Roncal-Herrero, H. Peirson, A.P. Brown, S.J. Milne, Temperature dependence of incommensurate modulation in $Ca_{0.28}Ba_{0.72}Nb_2O_6$, J. Appl. Phys. 134, 064101 (2023).

18  M. Eßer, M. Burianek, D. Klimmb, M. Muhlberg, Single crystal growth of the tetragonal tungsten bronze $Ca_xBa_{1-x}Nb_2O_6$, Journal of Crystal Growth 240, 1 (2002).

19  Š. Svirskas, D. Jablonskas, S. Rudys, S. Lapinskas, R. Grigalaitis, J. Banys, Broad-band measurements of dielectric permittivity in coaxial line using partially filled circular waveguide, *Rev. Sci. Instrum.* **91**, 035106 (2020).

20  J. Grigas, *Microwave Dielectric Spectroscopy of Ferroelectrics and Related Materials*, Gordon and Breach publishers, 1996.

21  F. Gervais, *Infrared and Millimeter waves*, vol. 8, Chapter 7, edited by K. J. Button, (Ac. Press, New York, 1983), p. 279.
22  E. Buixaderas, I. Gregora, J. Hlinka, J. Dec, T. Lukasiewicz, Raman and IR phonons in ferroelectric $Sr_{0.35}Ba_{0.69}Nb_2O_{6.04}$ single crystals, *Phase Transitions*, **86**, 217 (2013).
23  W. Hayes, R. Loudon, *Scattering of Light by Crystals*, John Wiley, New York, 1978.
24  J. Alanis. M.C. Rodríguez-Aranda. Á.G. Rodríguez *et al*. Temperature dependence of the Raman dispersion of $Sr_2Nb_2O_7$, *J. Raman Spectrosc.* **50**, 102 (2019).
25  E. Buixaderas, S. Kamba and J. Petzelt, Polar phonons and far-infrared amplitudon in $Sr_2Nb_2O_7$, *J. Phys. Condens. Matt.* **13**, 2823 (2001).
26  I. Levin, E. Cockayne, M.W. Lufaso, J.C. Woicik, J.E. Maslar, Local Structures and Raman Spectra in the $Ca(Zr,Ti)O_3$ Perovskite Solid Solutions, *Chem. Mater.* **18**, 854 (2006).
27  M. Aftabuzzaman, J. Dec, W. Kleemann, S. Kojima, Field dependent elastic anomaly in uniaxial tungsten bronze relaxors, *Jap. J. Appl. Phys.* **55**, 10TC01 (2016).
28  E. Buixaderas, V. Bovtun, M. Kempa, D. Nuzhnyy, M. Savinov, P. Vaněk, I. Gregora, and B. Malič, Lattice dynamics and domain wall oscillations of morphotropic $Pb(Zr,Ti)O_3$ ceramics, *Phys Rev. B* **94**, 054315 (2016).
29  G. Arlt, Ferroelastic domain walls as powerful shear wave emitters at microwaves, *Ferroelectrics* **172**, 95 (1995).
30  R. Paszkowski, K. Wokulska, T. Lukasiewicz, J. Dec, Cryst. Res. Technol. **48**, 413 (2013)
31  J. Dec, W. Kleemann, V. V. Shvartsman, D. C. Lupascu, T. Łukasiewicz, From mesoscopic to global polar order in the uniaxial relaxor ferroelectric $Sr_{0.8}Ba_{0.2}Nb_2O_6$, *Appl. Phys. Lett.* **100,** 052903 (2012).